\documentclass[twocolumn,showpacs,preprintnumbers,amsmath,amssymb]{revtex4}
\usepackage{graphicx}
\begin{document}
\title {Time-evolution stability of order parameters and phase
diagrams of bosons on optical lattice}
\author{Yue Yu}
\affiliation{Institute of Theoretical Physics, Chinese Academy of
Sciences, P.O. Box 2735, Beijing 100080, China}
\begin{abstract}
Stemming from the Heisenberg equations of motion, we study the
time-evolution stability of the order parameters for the cold
atoms on optical lattices. The requirement of this stability of
the order parameters endows the phase diagram with a fruitful
structure in the superfluid phase.  For the one-component
Bose-Hubbard model, we see that this stability of order parameter
leads to a physically receivable phase diagram. For two-component
bosons, we show that the molecules are preformed in the atomic
superfluid and then condenses into a molecular superfluid phase at
a critical repulsive inter-species interaction, which resembles
the pre-pairing mechanism in high $T_c$ superconductor of Cu-O
cuprates.
\end{abstract}


\maketitle

Ultra-cold atoms on optical lattices have offered a paradigmatic
scene to study strongly correlated phenomenon in a highly
controllable environment\cite{1,2,3,kohl}.
Superfluid/Mott-insulator transition was witnessed on $^{87}$Rb
atoms in a three- \cite{2} and one-dimensional lattice
\cite{kohl}. Theoretically, the strongly correlated phenomenon for
boson systems may be studied by the Bose-Hubbard model \cite{BH},
and experimental feasibility was also established by microscopic
calculations of the model parameters for cold boson atoms in
optical lattices \cite{jaks}.

Very fruitful structures  of the phase diagram have been explored
. The phase boundary between the superfluid and the Mott insulator
has been determined \cite{sheshadri} and verified by the
experiments \cite{2,kohl}. In the Mott insulator, the Bose-Hubbard
model can be mapped into an effective spin Hamiltonian and the
phase structure has been extensively studied \cite{kul}. On the
superfluid side, it is also predicted that there are complex
quantum phase transitions \cite{dem,kul}.

The Bose-Hubbard  model can only be exactly solved in
one-dimension for a limited case. Various approximations have been
applied. If an approximation ground state describes a spontaneous
breaking of symmetry, it has to be stable in the time-evolution. A
powerful tool to study this stability is the Heisenberg equations
of motion of the corresponding order parameter operator. For any
equilibrium state described by a time-independent density matrix
$\rho$ with $[H, \rho]=0$, the Heisenberg equation of motion of an
operator $O$ yields
\begin{eqnarray}
-i\langle \dot O\rangle= \langle[H,O]\rangle={\rm Tr}(\rho
[H,O])=0,\label{hem}
\end{eqnarray}
which gives the exact constraint on order parameter: When an
approximation state is readily to describe an ordered state, one
has to check that the corresponding order parameter is stable or
not in the sense obeying (\ref{hem}). The first such example has
been provided by Zhang for the Hubbard model of the electrons and
it has been generalized to the $t$-$J$ model \cite{zhang}. These
works suggested that the Hubbard model may not be a good model for
the high $T_c$ superconductor while the $t$-$J$ model is possible.

For bosons on the lattice, we shall show that similar exact
constraints on the order parameters play an important role in the
study of the phase structures of the systems. We shall use the
variational wave function which yields the mean field state for
Bose-Hubbard systems \cite{rk,dem}. It is seen that the
variational energy will be affected by the constraint on the
corresponding order parameter. We shall see that even for the
simplest one-component Bose-Hubbard model, these exact results
lead to a more physically receivable phase diagram. For the
two-component Bose-Hubbard model with a commensurate filling, the
most intriguing finding is that except the up-critical point from
the superfluid to the Mott insulator, there is a phase transition
from the atomic superfluid (AS) to the molecular superfluid(MS) in
a low-critical point for a repulsive inter-species interaction.
The existence of the MS phase for $U>0$ is because the local
$n_i=1$ is violated in the superfluid and $V>U$ favors two atoms
in different species to form a molecule. Our analysis will show
that the molecules are formed in the AS phase and condensed as $U$
reduces to a critical value, which resembles the prepairing
mechanism for the high T$_c$ superconductivity. To be specific, we
confine our discussions to the single atom filling per lattice
site in this work. However, it is easy to be generalized to other
integer
filling factors and the mixture of the Bose-Fermi atoms.\\

{\noindent \it One-component model:} We now begin with the
one-component Bose-Hubbard model, which is described by the
Hamiltonian
\begin{eqnarray}
H_{1}=-t\sum_{\langle ij\rangle}(a_i^\dag a_j+{\rm
h.c.})+\frac{V}{2}\sum_i n_i(n_i-1)-\mu\sum_i n_i,\label{sbhh}
\end{eqnarray}
where the notation is standard. For the uniform order parameters,
the Heisenberg equations of motion for $a$ and $a^2$ give
\begin{eqnarray}
-i\langle \dot a\rangle &=&tz\langle a\rangle -V\langle na\rangle
+\mu \langle a\rangle,\nonumber\\
-i\langle \dot {a^2}\rangle&=&2tz \langle a_{\delta}a\rangle+2\mu
\langle a^2\rangle-V\langle na^2\rangle,\label{sid}
\end{eqnarray}

where $a_\delta$ is the boson operator at a nearest neighbor and
$z$ is the partition number. Precisely, the superfluid and
dimerized order parameters are defined by $\langle a\rangle={\rm
Tr}(\rho a)$ and $\langle a^2\rangle={\rm Tr}(\rho a^2)$. In the
zero temperature limit, they are reduced to the expectation value
corresponding to the ground state. In the dilute gas limit,  a
widely accepted approxiamtion is, instead of the exact ground
state, to introduce a site factorizable wave function which takes
the form \cite{rk}
\begin{eqnarray}
|\Phi_1\rangle=\prod_i\biggl[\sin\frac{\theta_i}2 a_i^\dag
+\cos\frac{\theta_i}2\biggl( \sin\frac{\chi_i}2
+\frac{1}{\sqrt{2}} \cos\frac{\chi_i}2 a_i^{\dag
2}\biggr)\biggr]|0\rangle.
\end{eqnarray}
We confine the filling number per site to be one throughout this
work. The order parameters now are $\langle a\rangle=(2+\sqrt
2)\sin \theta/(2\sqrt 2)$ and $\langle
a^2\rangle=\cos^2\frac{\theta}2/\sqrt 2$ since $\chi=\pi/2$ when
the order parameters do not vanish and $\langle n\rangle =1$. If
we do not consider the time evolution of the order parameters,
there are three possible phase: the Mott insulator phase with
$\theta=\pi$ ($\langle a\rangle=\langle a^2\rangle=0$), the
dimerized phase with $\theta=0$ ($\langle a\rangle=0$ and $\langle
a^2\rangle\ne0$) and the superfluid phase with both $\langle
a\rangle\ne0$ and $\langle a^2 \rangle\ne 0$. If taking the
chemical potential $\mu$ as a constant, the previous variational
result has shown the critical points of the SF/MI and SF/D are in
$V/zt\approx\pm 5.83$ \cite{sheshadri}. The plus one is reasonable
and has been examined experimentally. The minus one seems to be a
mean field artifact because it was known that in the
one-dimensional model, the dimerized state dominates when
$V/zt\leq -1$ for $V<0$ \cite{hald}. To improve the result from
this simple mean field theory, we consider the stability of the
order parameters in time evolution. There are now two different
variational states in superfluid phase. One is that $\langle
a\rangle$ is stable and $\langle a^2\rangle$ is not and another is
vice verse. The former subjects to the constraint $-i\langle \dot
a\rangle=0$ which turns the first equation of (\ref{sid}) to $
\mu=\mu_1=-tz+V({\sqrt 2-1}). $ The latter to $-i \langle
a^2\rangle=0$, which turns to $ \mu=\mu_2=-\frac{3+2\sqrt 2}{\sqrt
2}tz\sin^2\frac{\theta}2. $ The variational energies to be
minimized are
\begin{eqnarray}
E_{s1,s2}(\theta)=-\frac{3+2\sqrt
2}8tz\sin^2\theta+\frac{V}2\cos^2\frac{\theta}2-\mu_{1,2}.
\end{eqnarray}
Minimizing these variational energies, the optimizing parameter is
given by
\begin{eqnarray}
\cos\theta_1=-\frac{V}{tz(3+2\sqrt{2})},~~
\cos\theta_2=\sqrt{2}-\frac{V}{tz(3+2\sqrt{2})}.
\end{eqnarray}
The requirement $|\cos\theta_{1,2}|\leq 1$ gives the regimes of
the minimal $E_{1s,2s}$ defined: $V^{(-)}_{1,2}\leq V\leq
V^{(+)}_{1,2}$ with $V_1^{(\pm)}
 /zt\approx\pm 5.83$ and $V_2^{(\pm)}/zt\approx 5.83(\sqrt 2\pm
 1)$. Outside of these ranges, the energies are defined by
 $E_{1,2}(\theta=0,\pi)$. Thus, corresponding to two cases,
the optimizing energies are
\begin{eqnarray}
E_{1,2}&=&E_{s1,s2}(\theta=0)\Theta(V_{1,2}^{(-)}-V)\nonumber
\\&+&E_{s1,s2}(V/tz)
\Theta(V-V_{1,2}^{(-)})\Theta(V_{1,2}^{(+)}-V)
\nonumber\\&+&E_{s1,s2}(\theta=\pi)
\Theta(V-V_{1,2}^{(+)}),\label{sve}
\end{eqnarray}
where $\Theta(x)$ is the step function. We depict these energies
in Fig. 1(a). It is found that in the whole range where $\langle
a\rangle\ne0$, $E_2>E_1$. However, at the point $V/zt\approx
-0.4$, $E_2=E_1$. This gives that the lower critical point of SF/D
transition. The whole phase diagram of the system is shown in Fig.
1(b). There are three phases which are identical to those in Fig.
1 (b). The phase with stable $\langle a^2\rangle\ne 0$ and nonzero
but unstable $\langle a\rangle$ does not exist. However, due to
the energy of the dimerized phase is lowered, the phase boundary
between SF-D and D is shifted a much reasonable value $V/tz\approx
-0.4$.

In our variational state, the states with three or more atoms per
site have been neglected. For a repulsive interaction, this
approximation is appropriate because there was a large three-body
collision loss in experiment. For a large negative $V$ or denser
atoms, the order parameters $\langle a^n\rangle$ ($n\geq 3$) are
possibly stable. We can use the procedure provided before to
discuss. Our procedure can also be applied to the other integer
filling by taking an appreciate
variational wave function.\\

{\noindent \it Two-component model: } Turn to the two-component
Bose-Hubbard model. The Hamiltonian is given by
\begin{eqnarray}
H_{2}&=&-\sum_{\langle
ij\rangle;\sigma=\uparrow,\downarrow}t_\sigma(a_{i\sigma}^\dag
a_{j\sigma}+{\rm h.c.})+\frac{1}2\sum_{i,\sigma}V_\sigma
n_{i\sigma}(n_{i\sigma}-1)\nonumber\\&+&U\sum_i(n_{i\uparrow}-\frac{1}2)
(n_{i\downarrow}-\frac{1}2)-\sum_{i\sigma}\mu_\sigma n_{i\sigma}.
\label{dbhh}
\end{eqnarray}
Generally, the tunneling amplitudes $t_{\uparrow,\downarrow}$, the
intra-species interactions $V_{\uparrow,\downarrow}$ may be
different. It was known that if $V_{\uparrow}V_{\downarrow}<U^2$,
one can not minimize the ground state energy with two nonzero
occupation numbers $\langle n_{\uparrow,\downarrow}\rangle$. The
system must be phase separated \cite{ps}. We shall not discuss the
phase separation here and consider only the interactions
restricted in $V_{\sigma}>0$ and $V_{\uparrow}V_{\downarrow}>U^2$
if $U>0$. For $U<0$, the phase diagram has been extensively
studied and we shall not go to the details here \cite{bcs,kul}.
The total atom number is conserved in the system and the ratio
between the different species atom numbers is self-consistently
determined by the internal field and the constraints given rise to
by the Heisenberg equations of motion. We now have five possible
order parameters, $\langle a_\sigma\rangle$, $\langle a_\uparrow
a_ \downarrow \rangle$ and $\langle a^2_\sigma\rangle$ if we fix
the filling factor per site to one and neglect the three or more
atom occupation in a single site. The time-evolutions of the order
parameters are given by
\begin{eqnarray}
&&-i\langle \dot
a_\sigma\rangle=(zt_\sigma+\mu_\sigma+\frac{U}2)\langle
a_\sigma\rangle-U\langle n_{\bar\sigma}a_\sigma\rangle-V_\sigma
\langle n_\sigma a_\sigma\rangle,\nonumber\\
&&-i\langle\dot
{a^2}_\sigma\rangle=(2zt_\sigma+2\mu_\sigma+U)\langle
a_{\delta,\sigma}a_\sigma\rangle-2U\langle
n_{\bar\sigma}a^2_\sigma\rangle\nonumber\\
&&~~~~~~~~~~~~~~-V_\sigma \langle n_\sigma
a^2_\sigma\rangle,\nonumber\\
&&-i\langle \dot{(a_\uparrow a_\downarrow)}
\rangle=\frac{1}2z(t_\uparrow+t_\downarrow)(\langle
a_{\delta,\uparrow} a_\downarrow \rangle+\langle
a_{\delta,\downarrow} a_\uparrow \rangle)\nonumber\\
&&~~~~~~~~~~~~~~+(\frac{V}2-U)\langle n a_\uparrow a_\downarrow
\rangle+(\mu_\uparrow+\mu_\downarrow)\langle a_\uparrow
a_\downarrow \rangle. \label{id}
\end{eqnarray}
The most general factorable variational wave function in our case
is that
\begin{eqnarray}
|\Phi_2\rangle&=&\prod_i\biggl[\sin\frac{\theta_i}2 \biggl
(\sin\frac{\chi_i}2a_{i\uparrow}^\dag
+\cos\frac{\chi_i}2a_{i\downarrow}^\dag\biggr)\nonumber\\
&+&\cos\frac{\theta_i}2\biggl( \sin\frac{\eta_i}2
+\cos\frac{\eta_i}2 \biggl(\sin\frac{\xi_i}2a_{i\uparrow}^\dag
a_{i\downarrow}^\dag\nonumber\\
&+&\frac{1}{\sqrt{2}}\biggl(\sin\frac{\zeta_i}2
a_{i\uparrow}^{\dag 2}+\cos\frac{\zeta_i}2a_{i\downarrow}^{\dag
2}\biggr)\biggr)\biggr)\biggr]|0\rangle. \label{wf}
\end{eqnarray}
Note that this factorable assumption implies that an approximation
neglecting all the spatial correlations. Thus, the BCS-type state
may not appear in our phase diagram although it is a possible
ground state \cite{bcs}. The condition $\langle n\rangle =1$ fixes
$\eta=\pi/2$ if $\theta\ne \pi$. The Mott insulator phase for this
Hamiltonian and the variational wave function have been
extensively studied by authors of Ref. \cite{kul}. We here focus
on the homogenous state although we can not exclude the
inhomogenous states such as spin glass and so no. They are the
atomic superfluid state with $\langle a_\sigma\rangle\ne 0$, the
molecular superfluid state with $\langle a_\uparrow
a_\downarrow\rangle\ne 0$ and the dimerized state with $\langle
a_\sigma^2\rangle\ne 0$.

In terms of (\ref{id}), the constraint identities may be obtained
by requiring the stability of these order parameters. The
variational energy now is given by
\begin{eqnarray}
E&=&-t_\uparrow z\langle a_\uparrow\rangle^2-t_\downarrow z\langle
a_\downarrow\rangle^2+U\langle n_\uparrow
n_\downarrow\rangle-\frac{U}4 \label{dv}\\&+&\frac{\langle
n^2_\uparrow\rangle+\langle n^2_\downarrow\rangle}2
-(V_\uparrow/2+\mu_\uparrow)\langle n_\uparrow
\rangle+(V_\downarrow/2+\mu_\downarrow)\langle n_\downarrow
\rangle,\nonumber
\end{eqnarray}
where $\mu_\sigma$ are restricted by the stability requirement of
the order parameters. All expectation values in (\ref{id}) and
(\ref{dv}) may be straightforwardly calculated according to the
variational wave function (\ref{wf}).

The phase diagram may obtained by solving the constraint equations
for the stable order parameter and minimizing the variational
energy (\ref{dv}). To save the space, we do not discuss strongly
asymmetric case with $t_\uparrow/V_\uparrow\ll
t_\downarrow/V_\downarrow$ in which one-component bosons may in
the Mott phase while the other is in the superfluid phase. We
focus on the species-symmetric model with
$t_\uparrow=t_\downarrow=t$ and $V_\uparrow=V_\downarrow=V$. The
symmetry requires that the order parameters are
species-independent and leads to $\mu_\uparrow=\mu_\downarrow=\mu$
and $\chi=\zeta=\pi/2$. The variational parameters are reduced to
$\theta$ and $\xi$ only. Corresponding to the stability of
non-zero $\langle a_\uparrow\rangle=\langle a_\downarrow\rangle$,
$\langle a^2_\uparrow\rangle=\langle a^2_\downarrow\rangle$, or
$\langle a_\uparrow a_\downarrow\rangle$, the constraint
identities are
\begin{eqnarray}
&&(zt+\mu+U/2)\biggl(1+\sin\frac{\xi}2+\cos\frac{\xi}2\biggr)
=U\biggl(\sin\frac{\xi}2+\cos\frac{\xi}2\biggr),\nonumber\\
&&zt\sin^2\frac{\theta}2
\biggl(1+\sin\frac{\xi}2+\cos\frac{\xi}2\biggr)^2 +2(\mu+U/2)\cos
\frac{\xi}2=0,\nonumber\\
&&zt\sin^2\frac{\theta}2
\biggl(1+\sin\frac{\xi}2+\cos\frac{\xi}2\biggr)^2 +2\mu\sin
\frac{\xi}2=0. \label{hcs}
\end{eqnarray}
And the variational energy reads
\begin{eqnarray}
E&=&-\frac{tz}8\sin^2\theta
\biggl(1+\sin\frac{\xi}2+\cos\frac{\xi}2\biggr)^2\label{dve}\\&+&\frac{1}2
\cos^2\frac{\theta}2\biggl(V\cos^2\frac{\xi}2+U\sin^2\frac{\xi}2\biggr)
-\mu-\frac{U-V}4.\nonumber
\end{eqnarray}
Two cases are analytically solvable. One is the hard-core limit in
which $V/U\to \infty$ and another is the critical point of the
phase separation, $U=V$. In the hard-core limit, the energies are
minimized at $\xi=\pi$. This gives $\langle a^2_\sigma\rangle=0$
as it should be in the hard-core limit. The atomic superfluid
phase (AS/MS) corresponding to the first constraint in (\ref{hcs})
is confined $-1<U/4tz<1$ while the molecular superfluid phase
(MS/AS) to the last constraint in (\ref{hcs}) is confined
$0<U/4tz<2$. The variational energies corresponding to these two
order parameters in the whole regime of the interaction $U$ may be
given by in a similar form as (\ref{sve}). Comparing these two
energies for any given $U/4zt$, the phase diagram in the hard-core
limit for the symmetric model can be determined (see Fig. 2(a)).
The four phases can detected: the molecular superfluid phase and
the mixture of the stable molecular superfluid and unstable atomic
superfluid phases, these two phases corresponds to the $s$-wave
superconductivity. The mixture of the stable atomic superfluid and
unstable molecular superfluid phases, which corresponds to the
preparing (but $d$-wave) for the electrons in high T$_c$ materials
; and the Mott insulator phase. The boundary of the second and
third ones is in $U/4zt=0.5$.

For any finite $U/V$, to keep the variational energy having a
lower bound, the second and third constraints in (\ref{hcs}) can
only be satisfied if $\theta=0$, and $\mu=-U/2$ and 0,
respectively. In the critical point of phase separation, $U=V$,
$E_2=3V/2$ for the second constraint and $E_3=V$ for the third one
and then the dimerized phase is always unfavorable. The energy
corresponding to the first constraint is minimized at $\xi=\pi/2$.
We compare $E_3$ and $E_1$ which are in a similar form as
(\ref{sve}) and the phase diagram can be depicted (see Fig. 2(b)).

For a general finite $U/V<1$, the variational energies may be
numerically minimized and the phase diagrams are plotted in Fig.
2(b) for $U>0$.  It is seen that the 'MS/AS-D' phase can only
exist in the hard-core limit. For a finite $V/U$, this phase is
suppressed and turns to the 'MS' phase. However, as $V$ reduces,
the atomic superfluid phase 'AS/MS-D' grows in both directions. In
our variational state, the suppression of the Mott phase is not so
strong when $V/U\leq 1.73$.

We now discuss the asymmetric case in the hard-core limit.  For
$t_\uparrow\ne t_\downarrow$, we find that the atomic superfluid
state with a stable $\langle a_\sigma\rangle$ exists for $
-\frac{1}2-\frac{1}{2\sin \chi_0}+\frac{\Delta
t\tan\chi_0}2<\frac{U}
{2z(t_\uparrow+t_\downarrow)}<\frac{1}2+\frac{1}{2\sin
\chi_0}+\frac{\Delta t\tan\chi_0}2, $ with $\chi_0$ is the
solution of the equation $\frac{1}{4(1+\sin\chi)}+\frac{U\sin\chi}
{2z(t_\uparrow+t_\downarrow)(1+\sin\chi)}+\frac{\Delta t\cos\chi}
{2(1+\sin\chi)}+\frac{2U} {z(t_\uparrow+t_\downarrow)}=\Delta
t\tan\chi$ and $\Delta
t=\frac{t_\uparrow-t_\downarrow}{t_\uparrow+t_\downarrow}$.
Solving the equation of $\chi$, one finds that there is only a
minor modification to that uses the simple variational method
taking $\mu_\sigma=0$. The latter gives
$|U|/2z(t_\uparrow+t_\downarrow)<1$ \cite{kul} and the former,
even for $\Delta t=\pm 1$, $|U|/2z(t_\uparrow+t_\downarrow)<0.97$.
At the other critical points, one also finds the critical values
of $|U|/2z(t_\uparrow+t_\downarrow)$ does not leave far from their
symmetric ones. Thus, we can think the basic feature of the phase
structure of superfluid phase in the hard-core limit can mostly be
described by the symmetric model. For finite $V_\sigma$, the
situation may be more complicated, e.g., for the strongly
asymmetric case, there is a mixture between the Mott insulator in
the one species and the superfluid in the other (see, \cite{kul},
Chen and Wu at \cite{ps}). We shall leave this for coming works.

Finally, we comment on the counterflow superfluid (see Kuklov et
al in  \cite{kul}). The corresponding order parameter is $\langle
a^\dag_\uparrow a_\downarrow\rangle=\frac{1}2[\sin^2\frac{\theta}2
\sin\chi+\sqrt{2}\cos^2\frac{\theta}2\sin\xi\cos^2
\frac{\eta}2(\sin\frac{\zeta}2+\cos\frac{\zeta}2)]$. For
$\theta=\pi$, the Mott state, $\langle a^\dag_\uparrow
a_\downarrow\rangle=(1/2)\sin\chi$. However, this mean field order
parameter is not stable because the Heisenberg equation of motion
 $ -i\langle \dot{(a^\dag_\uparrow
a_\downarrow)}\rangle=V(\langle
(n_\uparrow-n_\downarrow)a^\dag_\uparrow
a_\downarrow\rangle)=V\langle a^\dag_\uparrow a_\downarrow\rangle
$ for the Mott state. In the symmetric model, it holds for
arbitrary $U$ and $V0$. Thus, from a mean field point of view, the
counterflow superfluid is stable only in a time scale $t\ll 1/V$.

This work was supported in part by the National Science Foundation
of China.

{\noindent Fig. 1 (a) The variational energies associated with
different stable order parameters for the one-component
Bose-Hubbard model. (b) The corresponding phase diagram. 'D' is
the dimerized phase with $\langle a^2\rangle=1/2$ and $\langle
a\rangle=0$; 'SF-D' the superfluid phase with a stable $\langle
a\rangle\ne 0$ and unstable $\langle a^2\rangle\ne 0$; 'Mott' is
the Mott insulator phase with $\langle a\rangle=\langle
a^2\rangle=0$.}

{\noindent Fig. 2 The phase diagrams for different $V/U$. 'MS'
denotes the molecular superfluid phase with a stable $\langle
a_\uparrow a_\downarrow\rangle=\frac{1}2$ and other order
parameters vanish. 'MS/AS-D' denotes all order parameters
non-vanishing but only $\langle a_\uparrow a_\downarrow\rangle$
stable. 'AS/MS-D' does similarly but $\langle a_\sigma\rangle$
stable. (a) The hard-core limit. (b) Finite V/U.}


\begin{thebibliography}{a}
\bibitem{1} C. Orzel, A. K. Tuchman, M. L. Fenselau, M. Yasuda and M. A. Kasevich,
Science \textbf{291}, 2386 (2001).
\bibitem{2} M. Greiner, O. Mandel, T. Esslinger, T. W. H\"{a}nsch and I. Bloch,
Nature \textbf{415}, 39 (2002).

\bibitem{3} M. Greiner, I. Bloch, O. Mandel,
T. W. H\"{a}nsch and T. Esslinger, Phys. Rev. Lett. \textbf{87},
160405 (2001).

\bibitem{kohl} T. St\"oferle, H. Moritz, C. Schori, M. K\"ohl,
T. Esslinger, cond-mat/0312440.

\bibitem{BH} M. P. A. Fisher, P. B. Weichman, G. Grinstein, and D.
S. Fisher, Phys. Rev. B {\bf 40}, 546 (1989).

\bibitem{jaks} D. Jaksch, C. Bruder, J. I. Cirac, C. W. Gardiner
and P. Zoller,
Phys. Rev. Lett. \textbf{81}, 3108 (1998).

\bibitem{sheshadri} W. Krauth, M. Caffarel, and J.-P. Bouchard, Phys. Rev. B 45, 3137 (1992);
K. Sheshadri et al., Europhys. Lett. 22, 257 (1993); J. K.
Freericks and H. Monien, Europhys. Lett. 26, 545 (1994).

\bibitem{hald} T. C. Choy and F. D. M. Haldane, Phys. Lett. 90{\bf A}, 83
(1982).

\bibitem{kul} A. B. Kuklov and Svistunov, Phys. Rev, Lett. {\bf
90}, 100401 (2003); L. M. Duan, E. Demler and M. D. Lukin, Phys.
Rev. Lett. {\bf 91}, 090402 (2003); E. Altman, W. Hosfstetter, E.
Demler and M. D. Lukin, New J. Phys. {\bf 5}, 113 (2003). A.
Kuklov, N. Prokof'ev, and B. Svistunov Phys. Rev. Lett. {\bf 92},
030403 (2004); idib, 050402 (2004).


\bibitem{dem} E. Demler and F. Zhou, Phys. Rev. Lett. {\bf 88},
163001 (2002).

\bibitem{zhang} S. C. Zhang, Phys. Rev. B {\bf 42}, 1012 (1990).
Yue Yu and Zhongyuan Zhu, Phys. Rev. B {\bf 45}, 12502 (1992).

\bibitem{rk} D. S. Rokhsar and B. G. Kotliar, Phys. Rev. B {\bf
44}, 10328 (1991).

\bibitem{ps} J. P. Burke, et al, Phys. Rev. A {\bf 55}, 2511
(1997); H. m. J. M. Boesten et al, Phys. Rev. A {\bf 55}, 636
(1997); P. Ao and S. T. Chui, Phys. Rev. A {\bf 58}, 4836 (1998).
G. H. Chen and Y. S. Wu, Phys. Rev. A {\bf 67}, 013606 (2003).

\bibitem{bcs} M. Yu. Kagan and D. V. Efremov, Phys. Rev. B {\bf 65},
 195103 (2002); B. Paredes and J. I. Cirac, Phys. Rev. Lett. {\bf
 90},  150402 (2003).

\end{thebibliography}
\end{document}